\input{aipcheck}

\documentclass[
    ,final            
  ]
  {aipproc}

\layoutstyle{6x9}
\usepackage{amssymb}

\begin{document}

\title{A Mixed $\tau$-Electroproduction Sumrule for $\vert V_{us}\vert$}



\author{Kim Maltman}{
  address={Math and Stats, York Univ., 4700 Keele St., Toronto, ON CANADA
M3J 1P3}
  ,altaddress={CSSM, Univ. of Adelaide, Adelaide, 5005 SA, Australia} 
}

\begin{abstract}
A sum rule for determining $\vert V_{us}\vert$ from a combination of hadronic
$\tau$ decay and electroproduction data is discussed. Indications of 
problems with analogous, purely $\tau$-decay-based analyses,
most likely associated with slow convergence of the relevant $D=2$
OPE series, are outlined, and advantages of the mixed sum rule
approach identified. With current data, one obtains
$\vert V_{us}\vert =0.2208(39)_{exp}(3)_{th}$, a determination
amenable to significant near-term reduction in experimental
error, and with the smallest theoretical error of all current 
$\vert V_{us}\vert$ determinations.

\end{abstract}

\maketitle



\begin{figure*}
  \begin{minipage}[t]{0.47\linewidth}
\rotatebox{270}{\mbox{
\includegraphics[width=0.85\textwidth]{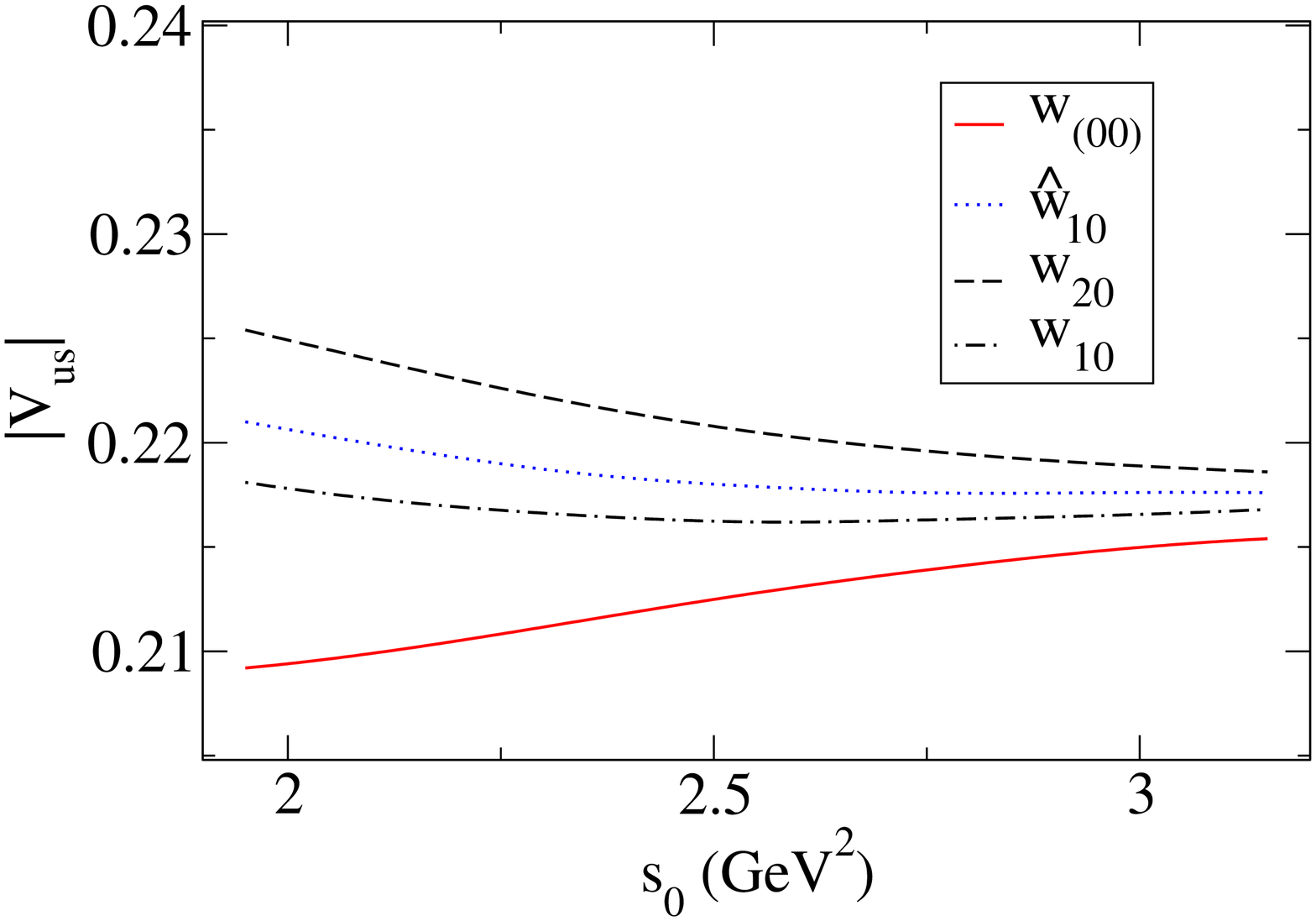}
}}
  \end{minipage}
\hfill
  \begin{minipage}[t]{0.47\linewidth}
\rotatebox{270}{\mbox{
\includegraphics[width=0.85\textwidth]{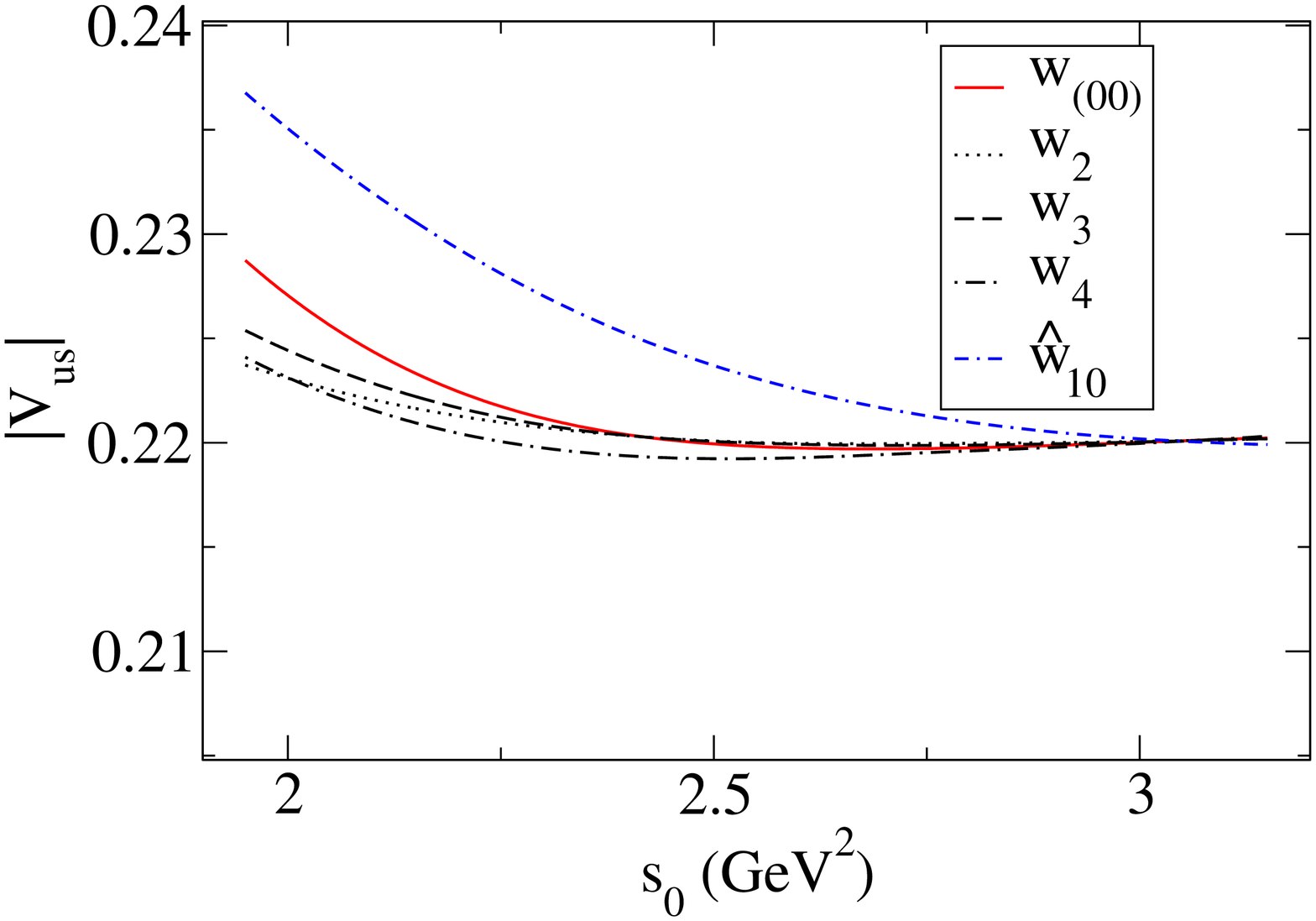}
}}
   \caption{$\vert V_{us}\vert$ versus $s_0$ from the 
$\Delta\Pi_\tau$ (left panel) and $\Delta\Pi_{\tau ,EM}$ (right panel)
FESRs for various $w(s)$. $w_{(00)}$ is the
conventional kinematic weight. See Ref.~\cite{kmtauem08} for
details of the other weights.
\label{vusfigure}}
\end{minipage}
\end{figure*}

In the SM, with $\Gamma^{had}_{V/A;ud}$ the $\tau$ width to hadrons through 
the flavor $ij$ V or A current, $\Gamma_e$ the $\tau$ electronic width, 
$y_\tau =s/m_\tau^2$, and $S_{EW}$ a known short-distance EW correction, 
$R_{V/A;ij}=\Gamma^{had}_{V/A;ij}/\Gamma_e$ is
related to the spectral functions $\rho^{(J)}_{V/A;ij}(s)$ of the
spin $J$ scalar correlators, $\Pi_{V/A;ij}^{(J)}(s)$, 
of the V/A current-current two-point functions by~\cite{tsai}
\begin{equation}
dR_{V/A;ij}/dy_\tau = 12\pi^2S_{EW} \vert V_{ij}\vert^2
\left[ w_{(00)}(y_\tau )\rho^{(0+1)}_{V/A;ij}(s)-w_L(y_\tau )
\rho_{V/A;ij}^{(0)}(s)
\right] ,
\label{taubasic}\end{equation}
where $w_{(00)}(y)=(1-y)^2(1+2y)$, $w_L(y)=2y(1-y)^2$ and, apart
from the $\pi$ and $K$ pole terms, $\rho^{(0)}_{V/A;ij}(s)$ are 
$O\left[ (m_j\mp m_i)^2\right]$. 
The basic finite energy sum rule (FESR) relation for the
correlators associated with the spectral function combinations 
in Eq.~(\ref{taubasic}), generically
\begin{equation}
\int_0^{s_0}w(s)\, \rho (s)\, ds\, =\, -{\frac{1}{2\pi i}}
\oint_{\vert s\vert =s_0}w(s)\, \Pi (s)\, ds\ ,
\label{basicfesr}
\end{equation}
and valid for all $s_0$, and any analytic $w(s)$, yields an equivalent 
contour integral representation for $R_{V/A;ij}$, one that can be turned 
into an OPE representation through the use of the OPE representation of 
$\Pi (s)$, for sufficiently large $s_0$, on the RHS of
Eq.~(\ref{basicfesr})~\cite{bnp}. This OPE representation
turns out to be extremely badly behaved for the purely
$J=0$ component~\cite{longprobs}, necessitating a focus
on the $J=0+1$ component alone. This is possible
because (i) $J=0$ contributions to $dR_{V/A;ij}/dy_\tau$
are dominated by the accurately known $\pi$ or $K$
pole terms, if present, and (ii) remaining small, but
not-completely-negligible, $O\left[ (m_s\mp m_u)^2\right]$-suppressed
$ij=us$, $J=0$ contributions can be estimated with sufficient
accuracy using sum rule and dispersive constraints~\cite{kmjk,jop}.
Defining the resulting $J=0+1$, $w(s)$-weighted spectral integrals
by
\begin{equation}
R^w_{V/A;ij}(s_0)\, \equiv\, 12\pi^2 S_{EW} \vert V_{ij}\vert^2\,
\int_0^{s_0}{\frac{ds}{m_\tau^2}}\, w(s)\, \rho^{(0+1)}_{V+A;ij}(s) \ ,
\end{equation}
one may form flavor-breaking differences
\begin{equation}
\delta R^w_{V+A}(s_0)\, =\, 
\left[ R^w_{V+A;ud}(s_0)/\vert V_{ud}\vert^2\right]
\, -\, \left[ R^w_{V+A;us}(s_0)/\vert V_{us}\vert^2\right]\ ,
\label{tauvusbasicidea}\end{equation}
whose OPE representation begins with a $D=2$ contribution $\propto m_s^2$ 
which, for typical $w(s)$, is {\it much}
smaller than the corresponding $R^w_{V+A;ud,us}(s_0)$.
Taking $\vert V_{ud}\vert$ and OPE input from elsewhere one
obtains the conventional pure-$\tau$-based
$\vert V_{us}\vert$ determination~\cite{gamizetalvus,kmcwvus}
\begin{equation}
\vert V_{us}\vert \, =\, \sqrt{{\frac{R^w_{V+A;us}(s_0)}
{{\frac{R^w_{V+A;ud}(s_0)}{\vert V_{ud}\vert^2}}
\, -\, \left[\delta R^w_{V+A}(s_0)\right]_{OPE}}}}\ ,
\label{tauvussolution}\end{equation}
which has been argued to have OPE-induced uncertainties
as small as $0.0005$. A potential problem with this
estimate  results from the slow convergence of
the $D=2$ OPE series for the relevant correlator
difference, $\Delta\Pi_\tau =\Pi_{V+A;ud}^{(0+1)}-\Pi_{V+A;us}^{(0+1)}$,
which, in the $\overline{MS}$ scheme,
with $a=\alpha_{\ s}(Q^2)/\pi$, $M_s=m_s(Q^2)$, has the
form~\cite{bck05}
\begin{equation}
\left[\Delta\Pi_\tau (Q^2)\right]^{OPE}_{D=2}\, =\, {\frac{3}{2\pi^2}}\,
{\frac{M_s^2}{Q^2}} \left[ 1\, +\, {\frac{7}{3}} a\, +\, 
19.93 {a}^2 \, +\, 208.75 {a}^3
\, +\, \cdots \right]\ ,
\label{d2tauform}\end{equation}
a result which shows very slow convergence given that
$a\sim 0.1$ for $Q_{}^2\sim m_\tau^2$. Fortunately, cross-checks
are possible. Not only can $\vert V_{us}\vert$ be
evaluated using different $s_0$ and $w(s)$, but
the integrated $D=2$ series can be evaluated
(i) using the truncated version of either 
$\left[ \Delta\Pi_\tau\right]^{D=2}_{OPE}$ itself or
the corresponding Adler function, and (ii) using either 
the contour improved (CIPT) or fixed order (FOPT) prescription.
The latter two options yield four combinations which differ from one another
only through contributions of order higher than
the truncation order. Eq.~(\ref{tauvussolution}) should then yield
$\vert V_{us}\vert$ compatible within error
estimates for any sufficiently large $s_0$, any $w(s)$, and 
any of these four evaluation schemes. 
With current data~\cite{aleph05,renewdata}, however, 
one finds that this is not the case. While changing from the $O({a}^3)$ 
to estimated $O({a}^4)$ version of the
Adler function-plus-CIPT scheme does yield a small ($\sim 0.0003$)
estimate for the corresponding theoretical error component,
changing instead from the $O({a}^3)$ Adler function-plus-CIPT 
scheme to, e.g., 
the $\left[ \Delta\Pi_\tau\right]_{OPE}^{D=2}$-plus-FOPT 
evaluation yields a shift in $\vert V_{us}\vert$ of $\sim 0.0023$.
In addition, as shown in the left panel of Fig.~\ref{vusfigure} 
(see Refs.~\cite{kmtauem08,kmvustau08} for details,
including more information on the weights employed), 
(i) significant instability of $\vert V_{us}\vert$ with
respect to $s_0$ is seen, especially for the 
conventionally employed kinematic weight, $w_{(00)}$, and
(ii) even at the highest accessible scale, 
$s_0=m_\tau^2$, non-trivial ($\sim \pm 0.0014$) variation with
$w(s)$ is seen. While there is clear evidence of convergence
towards a common value, albeit at $s_0$ higher than the
kinematic endpoint $s_0=m_\tau^2$, and some of the $s_0$-instability
and $w(s)$-dependence might be attributable to problems 
with the existing low-statistics $dR_{V+A;us}/ds$ data,
it is hard to justify as sufficiently conservative
estimates of theoretical uncertainties below $\sim 0.0020$ 
until such time as, at least,
(i) the relative reliability of the CIPT and FOPT schemes, and 
(ii) the origins of the observed $s_0$-instability
and $w(s)$-dependence are further investigated
and better understood.

A solution to the larger-than-ideal current theoretical
uncertainty in the $\Delta\Pi_\tau$-based determination
of $\vert V_{us}\vert$ is to consider analogous FESRs based
on correlator differences 
with OPE contributions strongly suppressed already at the
correlator level. A useful flavor-breaking correlator combination
having this property is~\cite{kmtauem08}
\begin{eqnarray}
&&\Delta\Pi_{\tau ,EM}\equiv 9\Pi_{EM}\, -\, \Pi_{V+A;us}^{(0+1)}
\, -\, 5\, \Pi_{V;ud}^{(0+1)}\,
+\, \Pi_{A;ud}^{(0+1)}\ ,
\end{eqnarray}
where $\Pi_{EM}$ is the scalar EM current-current correlator.
$\Delta\Pi_{\tau ,EM}$ is constructed to make exactly zero the leading term
in the $D=2$ OPE series. One finds that, with this
choice, other terms in the $D=2$ series are also
significantly suppressed, explicitly,
\begin{equation}
\left[\Delta\Pi_{\tau ,EM} (Q^2)\right]_{OPE}^{D=2}\, =\, {\frac{3}{2\pi^2}}\,
{\frac{M_s^2}{Q^2}} \left[ 0\, +\, {\frac{1}{3}} {a}\, +\,
4.3839 {a}^2\, +\, 44.943 {a}^3 \, +\, \cdots \right]
\label{d2tauemform}
\end{equation}
which should be compared to Eq.~(\ref{d2tauform}). Fortuitously,
$D=4$ contributions are also suppressed for $\Delta\Pi_{\tau ,EM}$
relative to $\Delta\Pi_\tau$.
Explicitly,
\begin{eqnarray}
\left[\Delta\Pi_{\tau ,EM} \right]^{OPE}_{D=4}
&=&{\frac{2}{Q^4}}\left[ 
\langle m_\ell {\overline{\ell}}\ell\rangle
\, -\, \langle {m_s}{\overline{s}}s\rangle\ \right]\left( 0-{\frac{4}{3}}{a}
\, -\, {\frac{59}{6}}{a}^2\right)\label{d4tauemform}\\
\left[\Delta\Pi_\tau \right]^{OPE}_{D=4}
&=& {\frac{2}{Q^4}}\left[ \langle m_\ell {\overline{\ell}}\ell\rangle
\, -\, \langle {m_s}{\overline{s}}s\rangle\ \right]\left( 1\, -\, {a}
\, -\, {\frac{13}{3}}{a}^2\right)\ . 
\label{d4tauform}\end{eqnarray}
The $D=2$ and $4$ suppressions are, however, not the result of some
additional symmetry since, e.g., the $D=6$ contribution
in the vacuum saturation approximation is a factor of $9/2$ larger for 
$\Delta\Pi_{\tau ,EM}$ than for $\Delta\Pi_\tau$. 
This is not a problem for the $\vert V_{us}\vert$ determination
since, if one uses weights $w(y)$ with $y=s/s_0$, integrated
OPE contributions of $D=2k+2$ scale as $1/s_0^k$
allowing any problems with assumed input values for combined
$D\geq 6$ condensate combinations to be exposed through
a study of the $s_0$-dependence of the $\vert V_{us}\vert$
produced by the analogue of Eq.~(\ref{tauvussolution}) for FESRs based on 
$\Delta\Pi_{\tau ,EM}$~\cite{kmtauem08}. The right panel of 
Fig.~\ref{vusfigure} shows the results for $\vert V_{us}\vert$
for a number of such $\Delta\Pi_{\tau ,EM}$ FESRs. Results
are shown, corresponding to $\tau$, $ij=us$ and EM spectral
data current as of October 2008~\cite{renewdata}, 
for the conventional $\tau$ kinematic weight, $w_{00}$,
the weight $\hat{w}_{10}$ which produced the best $s_0$-stability
for the $\Delta\Pi_\tau$ FESR, but is expected to have
enhanced $D>4$ contributions in the $\Delta\Pi_{\tau, EM}$ case, 
and for a selection of the weights, 
$w_N(y)=1-{\frac{N}{N-1}}y+{\frac{y^N}{N-1}}$, employed
in the recent $\tau$ determination of ${\alpha}_s$~\cite{kmty08}. 
The latter are useful since they have only a single, numerically 
suppressed $D=2N+2$ contribution beyond $D=4$. 
As one sees from the figure, most of the weights produce excellent
$s_0$-stability plateaus, and even the less stable
$\hat{w}_{10}$ case converges nicely to the common value
obtained using the other $w(y)$ as $s_0\rightarrow m_\tau^2$. 
Updating the results
of the figure for subsequently posted strange $\tau$ branching
fraction results~\cite{renewdata}, one finds~\cite{kmvustau08}
\begin{equation}
\vert V_{us}\vert = 0.2208\, (39)_{exp}\, (3)_{th}\,
\end{equation}
where the experimental error is due roughly equally to 
uncertainties in the $\tau$ $us$ and residual EM spectral
integrals. The theory error includes the uncertainty
in the $J=0$ phenomenological subtraction, and is an extremely conservative
one~\cite{kmtauem08}, one, moreover, born out by the $s_0$-stability
and $w(y)$-independence at $s_0=m_\tau^2$ of the relevant results.
Experimental errors should be subject to significant
near-term improvement, making this method, with its very
small theoretical error, an attractive one for generating a
$\vert V_{us}\vert$ determination independent of the conventional
ones based on $K$ physics.

\begin{theacknowledgments}
The hospitality of the CSSM, University of Adelaide,  
the Theory Group at IHEP, Beijing, and the ongoing support of
the Natural Sciences and
Engineering Research Council of Canada are gratefully acknowledged.
\end{theacknowledgments}


\begin{thebibliography}{9}
\bibitem{tsai}Y.-S. Tsai, \emph{Phys. Rev.} \textbf{D4}, 2821--2837 (1971).
\bibitem{bnp}E. Braaten, S. Narison and A. Pich, \emph{Nucl. Phys.} 
\textbf{B373}, 581--612 (1992).
\bibitem{longprobs}K. Maltman, \emph{Phys. Rev.} \textbf{D58}, 093015 (1998);
K. Maltman and J. Kambor, \emph{Phys. Rev.} \textbf{D64}, 093014 (2001).
\bibitem{km00}J. Kambor and K. Maltman, \emph{Phys. Rev.} \textbf{D62}, 093023 
(2000).
\bibitem{kmjk}K. Maltman and J. Kambor, \emph{Phys. Rev.} \textbf{D65},
074013 (2002). 
\bibitem{jop}M. Jamin, J.A. Oller and A. Pich, \emph{Nucl. Phys.} 
\textbf{B587}, 331--362 (2000); {\it ibid.} \textbf{B622}, 279--308 (2002); 
\emph{Phys. Rev.} \textbf{D74}, 074009 (2006).

\bibitem{gamizetalvus}E. Gamiz {\it et al.}, \emph{JHEP} \textbf{0301},
060 (2003); \emph{Phys. Rev. Lett.} \textbf{94}, 011803 (2005); 
\emph{PoS KAON 2007}, 008 (2008); A. Pich, \emph{Nucl. Phys. Proc. Suppl.} 
\textbf{181-182}, 300--305 (2008) 300.
\bibitem{kmcwvus}K. Maltman and C.E. Wolfe, \emph{Phys. Lett.} \textbf{B639},
283--289 (2006); {\it ibid.} \textbf{B650}, 27--32 (2007);
K. Maltman {\it et al.}, \emph{Int. J. Mod. Phys.} \textbf{A23}, 3191--3195
(2008).
\bibitem{bck05}P.A. Baikov, K.G. Chetyrkin and J.H. Kuhn,
\emph{Phys. Rev. Lett.} \textbf{95}, 012003 (2005).

\bibitem{aleph05}S. Schael {\it et al.} (The ALEPH Collaboration),
\emph{Phys. Rep.} \textbf{421}, 191--284 (2005).
\bibitem{renewdata}Space precludes listing all of the new $\tau$
$us$ and EM spectral data employed. See Refs.~\cite{kmtauem08,kmvustau08} 
for details.

\bibitem{kmtauem08}K. Maltman, \emph{Phys. Lett.} \textbf{B672}, 257--263
(2009).
\bibitem{kmvustau08}K. Maltman {\it et al.}, \emph{Nucl. Phys. Proc.
Suppl.} \textbf{189}, 175--180 (2009).
\bibitem{kmty08}K. Maltman and T. Yavin, \emph{Phys. Rev.}
\textbf{D78}, 094020 (2008).




\end{thebibliography}
\end{document}